\documentclass[journal]{IEEEtran}
\usepackage{graphicx}
\usepackage{amsmath}
\usepackage{color}
\usepackage{nopageno}
\usepackage{multicol}

\begin{document}
\title{Effects of Quantum Diffraction on the Propagation of E-A-W Solitary Structure in Fermi Plasma}

\author{S. Sarkar, T. Ghosh and S. Chandra
\thanks{S. Sarkar is with the Department of Physics, National Institute of Technology Karnataka, Karnataka 575025, India. (e-mail: sarkarsoumya65@gmail.com)}
\thanks{T. Ghosh is with the Physics Department, Bidhannagar Govt.College, Kolkata-700064, India (e-mail: tamalghosh695@gmail.com). }
\thanks{S. Chandra is with Govt. General Degree College at Kushmandi, Dakshin Dinajpur, 733121, India. (e-mail: swarniv147@gmail.com)}}
\maketitle
\begin{abstract}
Plasma state of matter can be studied in various types of situations. These studies are of great interest in Astrophysical objects like galaxies, accretion disk, neutron stars, etc, and laboratory plasma as well. Different objects demand different approaches to investigate the dynamics of the plasma. The relativistic effects in the motion of electrons in Quantum Plasma highly affect the characteristics of the solitary structure of the wave with two-temperature electrons. In this paper, considering the quantum hydrodynamic (QHD) model a dispersion relation is derived, and using standard perturbation technique, a mathematical model (i.e. nonlinear Schrödinger Equation) is studied for a wave with relativistic and quantum effects in it. We study the analysis for different values of diffraction coefficient, streaming velocity, and other plasma parameters as well. We analyze the stable rogue wave structure using NLSE and run simulations of those solitary profiles and rogue waves. 
\end{abstract}
\thispagestyle{empty}
\pagestyle{empty}
\section{Introduction}
\IEEEPARstart{A}{} gas when heated with higher temperature gets ionized. To qualify for a state of matter an ionized gas should satisfy some certain conditions. The fourth state of matter is that the Plasma state. As the interest of this paper, we work with the electron acoustic waves (EAWs) of plasma. EAWs consist of two groups of electrons - hot electrons and cold electrons. The temperature is understood statistically here, rather than our normal concept of it. Hot electrons contribute to the restoring force and cold electrons contribute to the inertia. Ions are considered to be immobile concerning electrons’ motion as it is bulkier. So we analyze the governing equations for electrons’ motion. The phase velocity of EAWs is $v_p \approx 1.4v$ \cite{ref1} at a small amplitude, this phase velocity is higher than the thermal velocity of cold electrons and lower than that of hot electrons \cite{ref2}. The non-relativistic motion for two-electron systems in electron acoustic waves and relativistic motion for ion-electron systems in ion-acoustic mode has already been studied. Here in this paper, we study the electron acoustic wave solitary structure with relativistic and quantum effects. Now it sometimes happens that a quantum system can be described by a classical equation and in that case, potential energy is modified by an effective quantum potential term. Here in describing the electrons’ motion, we get such a quantum potential term which is called Bohm potential \cite{ref3}. If a function $\phi(x,t)$ can be written as
\begin{equation}
    \phi(x,t) = A(x,t)e^{\frac{is(x,t)}{\hbar}}
\end{equation}
Where A(x,t)is the real amplitude, and s(x,t) is the real phase factor. Now by using the Schrödinger Equation we can derive the equation for the phase factor which contains an additional potential term Bohm Potential which causes the diffraction.
Studies have been done on the electron-acoustic mode in space plasma \cite{ref4,ref5,ref6} and also laboratory plasma \cite{ref7,ref8,ref9}. Further nonlinear evolution of EAW was analyzed \cite{ref10,ref11,ref12,ref13,ref14,ref15} and also reported to be seen in the auroral region\cite{ref16,ref17,ref18,ref19} and magnetosphere of Earth’s atmosphere \cite{ref20,ref21,ref22,ref23,ref24}. In 2019, a team studied the electron acoustic wave beyond the Bohm potential and they added a quantum correction to Coulomb potential \cite{ref25}. In recent years, the study of quantum-plasma has gained popularity, and using the Quantum Hydrodynamic model \cite{ref26} the properties of quantum plasma have been studied in large scales. The solitary structure for two-temperature electrons has been studied by Masood \cite{ref27}. Relativistic study of plasma has been done in the case of space plasma \cite{ref28}, earth magnetosphere \cite{ref29}, Van-Allen radiation belt \cite{ref30}, laser-plasma interaction \cite{ref31}. In a paper \cite{ref32}, it is also claimed that the relativistic motion of plasma was there in the early universe.
A degenerate matter is a highly dense fermionic matter that is generally found in the center of a star or in a neutron star where due to huge pressure the object becomes highly dense and the kinetic energy of the fermions is so high to maintain Pauli’s Exclusion Principle. When the velocity of the fermions becomes so close to the velocity of light in a vacuum then this is called relativistic degenerate matter. Due to Pauli’s Exclusion Principle, they begin to fill the higher energy states, and ‘degeneracy pressure’ comes to play its role. But here, we consider weakly-relativistic cases for both cold and hot electrons and in this case, the pressure law is defined by Fermi Pressure which is proportional to the cube of the number density of electrons. As hot electrons have much greater velocity compared to cold electrons we can ignore the oscillation of those, thus we can ignore restoring force term in the momentum equation, on the other hand, we can ignore the inertia term for cold electrons as it is mobile compared to hot electrons. Now, because of analytical ease and efficiency, the quantum hydrodynamic (QHD) model is widely used to study the non-relativistic regime. For relativistic study in the quantum regime, we can use the relativistic Wigner function described in \cite{ref33}. Thus we can generalize non-relativistic one-dimensional QHD to a weakly relativistic situation. Using the model for one-dimension we study the dispersion relation for different values of the ratio between hot electron density and cold electron density, quantum diffraction coefficient. 
The organization of the dissertation is as follows. In Section II, we formulate the basic equations that we need for further calculations, in Section III, we analyze the theoretical background of dispersion relation and the result. In Section IV, we study the KDV equation and see the modifying solitary structure with different parameters in quantum plasma. In Section V, we study the rogue wave deriving from the NLSE. In Section VI, we show the static plots of the simulations of solitary profiles from the KDV equation and Rogue waves from the NLSE, and finally, we conclude in Section VII. 


\section{Basic Equations}\label{BE}
As we consider a weakly relativistic regime for plasma with two-temperature hot and cold electrons with streaming motion in x-direction we assume the plasma acts as Fermi gas at zero temperature and thus the pressure law\cite{ref34} becomes
\begin{equation}\label{eq1}
P_{j}=\frac{m_{j} V_{F j}^{2}}{3 n_{j 0}^{2}} n_{j}^{3} 
\end{equation}
where $j$ is the index to indicate hot and cold electron, $j = h$ is for indicating hot electrons and $j = c$ is for indicating cold electrons; $m$ indicates the mass of electron; $n$ indicates the number density of electrons, and $V_F=\sqrt{\frac{2K_B T_{Fe}}{m_j}}$ is the fermi speed. Additionally, we take $u_{0j}$  as streaming motion and $n_{0j}$ are the value of number density in equilibrium.

The QHD set of equations to describe the dynamics of plasma can be written as below
\begin{equation}\label{eq2}
\frac{\partial n_{h}}{\partial t}+\frac{\partial\left(n_{h} u_{h}\right)}{\partial x}=0
\end{equation}
\begin{equation}\label{eq3}
\frac{\partial n_{c}}{\partial t}+\frac{\partial\left(n_{c} u_{c}\right)}{\partial x}=0
\end{equation}
\begin{equation}\label{eq4}
0=\frac{e}{m_{e}} \frac{\partial \phi}{\partial x}-\frac{1}{m_en_{h}} \frac{\partial P_{h}}{\partial x}+\frac{\hbar^{2}}{2 m_{e}}{\gamma_{h}^{2}} \frac{\partial}{\partial x}\left[\frac{1}{\sqrt{n_{h}}} \frac{\partial^{2} \sqrt{n_{h}}}{\partial x^{2}}\right]
\end{equation}

\begin{multline}\label{eq5}
\left(\frac{\partial}{\partial t}+u_{c} \frac{\partial}{\partial x}\right) (u_{c}\gamma_{c}) =\frac{e}{m_{e}} \frac{\partial \phi}{\partial x}\\
-\frac{1}{m_{e}n_{c}} \frac{\partial P_{c}}{\partial x} +\frac{\hbar^{2}}{2 m_{e}^{2}\gamma_{c}^{2}} \frac{\partial}{\partial x}\left[\frac{1}{\sqrt{n_{c}}} \frac{\partial^{2} \sqrt{n_{c}}}{\partial x^{2}}\right]
\end{multline}

\begin{equation}\label{eq6}
\frac{\partial^{2} \phi}{\partial x^{2}}=4 \pi e \left(n_{c}+n_{h}-Z_{i} n_{i}\right)
\end{equation}

Equations (3), (4), (5), (6), and (7) represent the continuity equation for hot electron, continuity equation for cold electron, momentum equation for hot electron, momentum equation for cold electron, and Poisson equation. Here, $\gamma_j= \left(1-\frac{u_j^2}{c^2} \right)^{-\frac{1}{2}}$  is the relativistic effect for electrons, and $\hbar$ is called the reduced Planck’s constant. Note we take relativistic effect for both hot and cold electrons. In the momentum equations the relativistic term comes from the effective mass of the electrons due to high velocity in weakly relativistic regime. To make the calculation easier we follow some normalization schemes to make the physical quantities dimensionless. The scheme we use is
 $x \rightarrow x \omega_{j} / V_{F j}, t \rightarrow t \omega_{j}, \phi \rightarrow e \phi / 2 k_{B} T_{F j}, u_{j} \rightarrow u_{j} / V_{F j} , n_{j} \rightarrow n_{j} / n_{j 0}, \eta_{c} \rightarrow \eta_{c} \omega_{j} / m_{e} V_{F j}^{2}$
where $\omega_{j}=\sqrt{4 \pi e n_{c 0} e^{2}/m_{e}}$ is called the plasma frequency, $k_{b}$ called the Boltzmann Constant. After using this normalization scheme the simplified QHD set of equations become

\begin{equation}\label{eq7}
\frac{\partial n_{h}}{\partial t}+\frac{\partial\left(n_{h} u_{h}\right)}{\partial x}=0
\end{equation}
\begin{equation}\label{eq8}
\frac{\partial n_{c}}{\partial t}+\frac{\partial\left(n_{c} u_{c}\right)}{\partial x}=0
\end{equation}
\begin{equation}\label{eq9}
0=\frac{\partial \phi}{\partial x}-n_{h} \frac{\partial n_{h}}{\partial t}+\frac{H^{2}}{2\gamma_{h}^2} \frac{\partial}{\partial x}\left[\frac{1}{\sqrt{n_{h}}} \frac{\partial^{2} \sqrt{n_{h}}}{\partial x^{2}}\right]
\end{equation}

\begin{multline}\label{eq10}
\left(\frac{\partial}{\partial t}+u_{c} \frac{\partial}{\partial x}\right) (u_{c}\gamma_c)=\frac{\partial \phi}{\partial x}\\+
\frac{H^{2}}{2\gamma_{c}^2} \frac{\partial}{\partial x}\left[\frac{1}{\sqrt{n_{c}}} \frac{\partial^{2} \sqrt{n_{c}}}{\partial x^{2}}\right]
\end{multline}

\begin{equation}\label{eq11}
\frac{\partial^{2}\phi}{\partial x^{2}}=n_{c}+\frac{n_{h}}{\delta}-\frac{\delta_{i}}{\delta} n_{i}
\end{equation}
Where $H=\hbar \omega_{j} / 2 k_{B} T_{F j}$ is called the ‘Quantum Diffraction term’ which is proportional to the ratio of plasma energy to Fermi energy $K_B T_{Fe}$.

\section{Dispersion Characteristics }\label{DR}

We can see dispersion characteristics in plasma i.e. its wave number changes with the frequency. To investigate the dispersion relation we follow standard perturbation technique where we expand the field variables $n_{j}, u_j,$ and $\phi$ about their equilibrium values as follows
\begin{equation}\label{eq12}
\left[\begin{array}{c}
n_{j} \\
u_{j} \\
\phi
\end{array}\right]=\left[\begin{array}{c}
n_{0} \\
u_{0} \\
0
\end{array}\right]+\varepsilon\left[\begin{array}{c}
n_{j}^{(1)} \\
u_{j}^{(1)} \\
\phi^{(1)}
\end{array}\right]+\varepsilon^{2}\left[\begin{array}{c}
n_{j}^{(2)} \\
u_{j}^{(2)} \\
\phi^{(2)}
\end{array}\right]+\cdots
\end{equation}

Assuming the field variables $V$ varies as $V=V_0 e^{i(k x-\omega t)},  (V\in n,u,\phi)$. Here, $V_0$  is the values in equilibrium and $V_{j}^{1}, V_{j}^{2}$ are the first ordered and second ordered perturbed terms, and so on. $\epsilon$ is the smallness parameter. Now by substituting Eq. (13) in the governing equations and taking only the first ordered coefficients of $\epsilon$ we get the dispersion relation

\begin{equation}\label{eq13}
1=\frac{1}{(\omega-ku_{0c})^2\gamma_{3}-\frac{k^4H^2}{4}\gamma_{2c}} - \frac{1/\delta}{k^2+\frac{k^4H^2}{4}\gamma_{2h}}
\end{equation}
 
By simplifying the above equation we get the expression of $\omega$ in terms of wavenumber $k$

 \begin{equation}\label{eq14}
 \omega=k u_{0} \pm\left(\frac{1}{\gamma_3}\left[\frac{k^2+\frac{k^4H^2}{4}\gamma_{2h}}{1/\delta+k^2+\frac{k^4H^2}{4}\gamma_{2h}}+\frac{k^4H^2}{4}\gamma_{2h}\right]\right)^{\frac{1}{2}}
\end{equation}
For lower values of $k(k\rightarrow0)$ i.e. in higher wavelength range we get

\begin{equation}\label{eq15}
\omega \approx k\left(u_{0c}+\sqrt\frac{\delta}{\gamma_3}\right)
\end{equation}

So in long-wavelength range, the phase velocity becomes

 \begin{equation}\label{eq16}
V_P = \frac{\omega}{k} k\left(u_{0c}+\sqrt\frac{\delta}{\gamma_3}\right)
\end{equation}

\subsection{Results and discussions}
 
\begin{figure}[!t]
{	\centering
	\includegraphics[width=3in,angle=0]{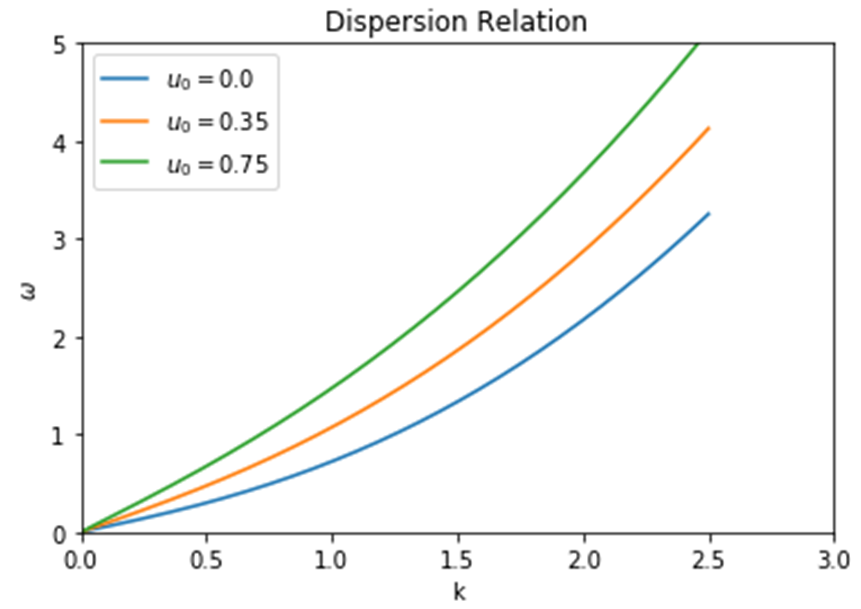}
	\caption{Dispersion curves for different values of streaming velocities keeping $H=1, \delta=0.3$}
	\label{fig1}
}
\end{figure}

\begin{figure}[!t]
{	\centering
	\includegraphics[width=3in,angle=0]{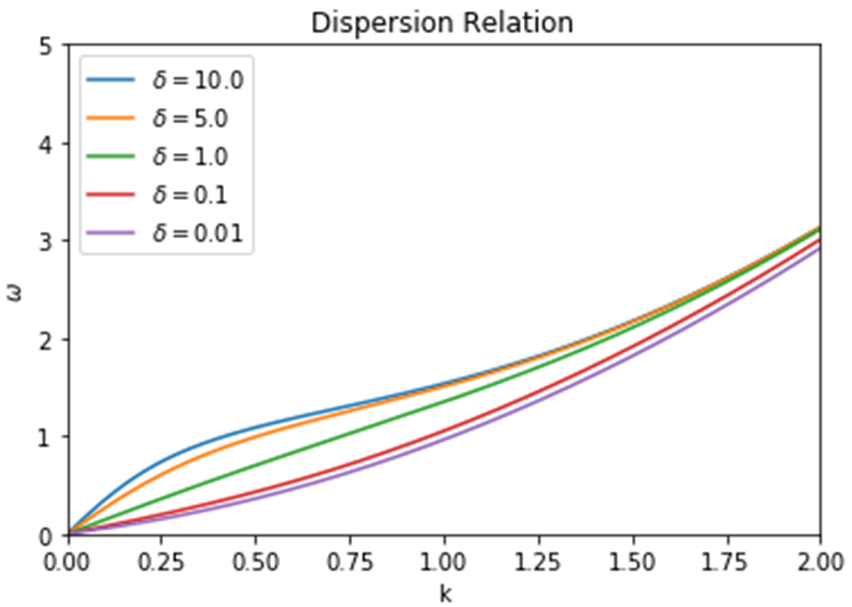}
	\caption{Dispersion curves for different values of $\delta$ keeping  $u_{0}=0.45, H=1$}
	\label{fig2}
}
\end{figure}

\begin{figure}[!t]
{	\centering
	\includegraphics[width=3in,angle=0]{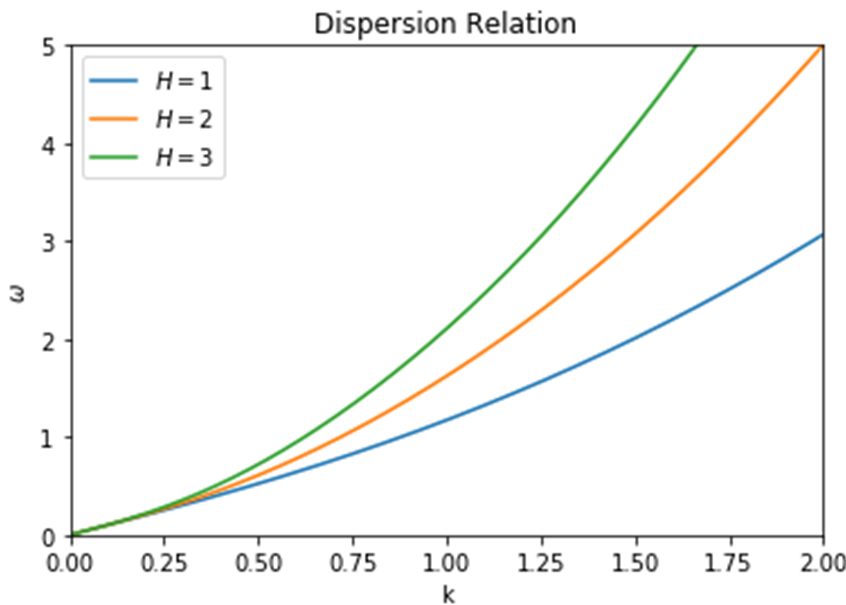}
	\caption{Dispersion relation for different values of H keeping $u_{0}=0.45, \delta=0.3$}
	\label{fig2}
}
\end{figure}

Using the quantum hydrodynamic model we have studied the effects of streaming motion, the ratio of hot-cold electron densities, and diffraction parameter in a relativistic quantum plasma analytically. Then we plot the effect of those parameters to understand the effects visually. In Fig. (1) we can see that the relativistic effect diminishes the nonlinear steepening of the relation. The steepening is always lesser for a non-zero streaming motion than that of zero streaming motion which is the non-relativistic case. Hence, for a particular value of k, with the increment of streaming value, the value of $\omega$ increases. For the higher value of streaming motion, the dispersion relation becomes almost linear. In Fig. (2) we have taken $\delta$ as $\delta=\frac{n_{0c}}{n_{0h}}$, so it's been clear from the figure when we are taking $n_{0c}\approx n_{0h}$, the dispersion relation is almost linear, both take part in the oscillation of the plasma. Otherwise, the relation shows nonlinear property. However, for large values of wave number $k$, the relation shows linearity anyway. In Fig. (3) it is seen that the diffraction parameter contributes to the nonlinear property of quantum plasma hugely. Another observation is that if we ignore the Bohm potential term and relativistic effects and also the equal distribution of cold and hot electron, we can see that at higher wave number $\omega$ attains a maximum value rapidly and continues with that.

\section{Derivation of Korteweg–de Vries equation}\label{KDV}
A solitary wave is a packet of waves that maintain its shapes while propagating with constant velocity. These solitons are created when dispersing force balances with the nonlinear force. To study the solitary profile of an electron acoustic wave we study the KDV equation. To do that we first follow a stretching of space and time

\begin{equation}\label{eq17}
\xi=\epsilon^{\frac{1}{2}}(x-M t) \;\;\;\;\;\text{and}\;\;\;\;\;  \tau=\epsilon^{\frac{3}{2}t}
\end{equation}
Where $M$ is the phase velocity in high range amplitude. Now using these stretching we replace the space and time coordinates in the equations (8)-(12) and equating the coefficients of the first order of the smallness parameter $\epsilon$ and get the number densities and velocities of two-temperature electrons in terms of potential

\begin{equation}\label{eq18}
n_{h}^{(1)}=Q_{1} \phi^{(1)} ; n_{c}^{(l)}=S_{1} \phi^{(1)} ; u_{h}^{(1)}=Q_{2} \phi^{(1)} ; u_{c}^{(1)}=R_{1} \phi^{(1)}
\end{equation}

After going for the second-order coefficient of the smallness parameter and linearizing we get the required \textbf{Korteweg–de Vries(KDV)} equation

\begin{equation}
\frac{\partial \phi^{(1)}}{\partial \tau}+A \phi \frac{\partial \phi^{(1)}}{\partial \xi}+B \frac{\partial^{3} \phi^{(1)}}{\partial \xi^{3}}=0 
\end{equation}
\begin{equation}
A=\frac{2 D S_{1} R_{1}+F n_{0 c} R_{1}^{2}-D\left(M-u_{0 c}\right) \frac{Q_{1}^{2}}{n_{0 h} \delta}}{D S_{1}+n_{0 c} R_{1}\left(1+\frac{3 u^{2} 0 c}{2 c^{2}}\right)}
\end{equation}
\begin{equation}
B=\frac{D\left(M-u_{0 c}\right) \frac{T_{h} Q_{1}}{n_{0} h^{\delta}}-T_{c} S_{1} n_{0 c}-D\left(M-u_{0 c}\right)}{D S_{1}+n_{0 c} R_{1}\left(1+\frac{3 u^{2} 0 c}{2 c^{2}}\right)}
\end{equation}

$A$ is called the non-linear coefficient and $B$ is called the dispersion coefficient. Now there are a lot of parameters up there in the equations (20), (21), and (22). To avoid the untidiness and for their frequent appearances we just define those parameters that look like

\begin{equation}
D=M+\frac{3 M u_{0 c}^{2}}{2 c^{2}}-\frac{3 u_{0 c}^{3}}{2 c^{2}}-u_{0 c}
\end{equation}
\begin{equation}
F=1+\frac{9 u_{0 c}^{3}}{2 c^{2}}-\frac{3 M u_{0 c}^{2}}{2 c^{2}} 
\end{equation}
\begin{equation}
T_{j}=\frac{H^{2}}{4 n_{0 j}^{2}}\left(1-\frac{u_{0 j}^{2}}{c^{2}}\right)
\end{equation}
\begin{equation}
S_{1}=\frac{n_{0 c}}{\left(M-u_{0 c}\right)\left[\left(u_{0 c}+\frac{3 u_{0 c}^{3}}{2 c^{2}}\right)-M\left(1+\frac{3 u_{0 c}^{3}}{2 c^{2}}\right)\right]}
\end{equation}
\begin{equation}
R_{1}=\frac{1}{\left(M-u_{0 c}\right)\left[\left(u_{0 c}+\frac{3 u_{0 c}^{3}}{2 c^{2}}\right)-M\left(1+\frac{3 u_{0 c}^{3}}{2 c^{2}}\right)\right]}
\end{equation}
\begin{equation}
Q_{1}=\frac{1}{n_{0 h}}
\end{equation}
\begin{equation}
Q_{2}=\frac{M-u_{0h}}{n_{0h}^2}
\end{equation}

To find the solution of Eq. (20) we follow another transformation of variables where we transform our $\xi$ and $\tau$ into one variable $\eta$ in a way like this: $\tau=\xi-m\tau$ , where $m$ is called the Mach number. Following the boundary conditions where at $\tau\rightarrow\pm\infty; \phi, \frac{\partial\phi}{\partial\eta},\frac{\partial^2\phi}{\partial\eta^2}\rightarrow0,$ we have our solution for Eq. (20) as follows

\begin{equation}
\phi=\phi_asec^2\left(\frac{\eta}{x}\right)
\end{equation}

Where $\phi_a$  is the amplitude and xis the width of the solitary wave and are given by

\begin{equation}
\phi_a=3\left(\frac{m}{A}\right)\;\;\;\;\;\text{and}\;\;\;\;\;x=\sqrt{\left(\frac{4B}{m}\right)}
\end{equation}

The relative strength of non-linear and dispersive force defines the structure of the solitary structure. As we can see from the Eq. (21) and (22), the relativistic effect changes the solitary structure. If we take $u_0=0$ and $n_{0j}=1$ , Eq. (21) and Eq. (22) reduce to

\begin{equation}
A=\frac{M^{3}}{2}-\frac{3}{2 M}
\end{equation}
\begin{equation}
B=\frac{H^{2}}{8 M}+\frac{M^{3}}{3}-\frac{M^{3} H^{2}}{4}
\end{equation}

\subsection{Results and discussions}
\begin{figure}[!htb]
{	\centering
	\includegraphics[width=3in,angle=0]{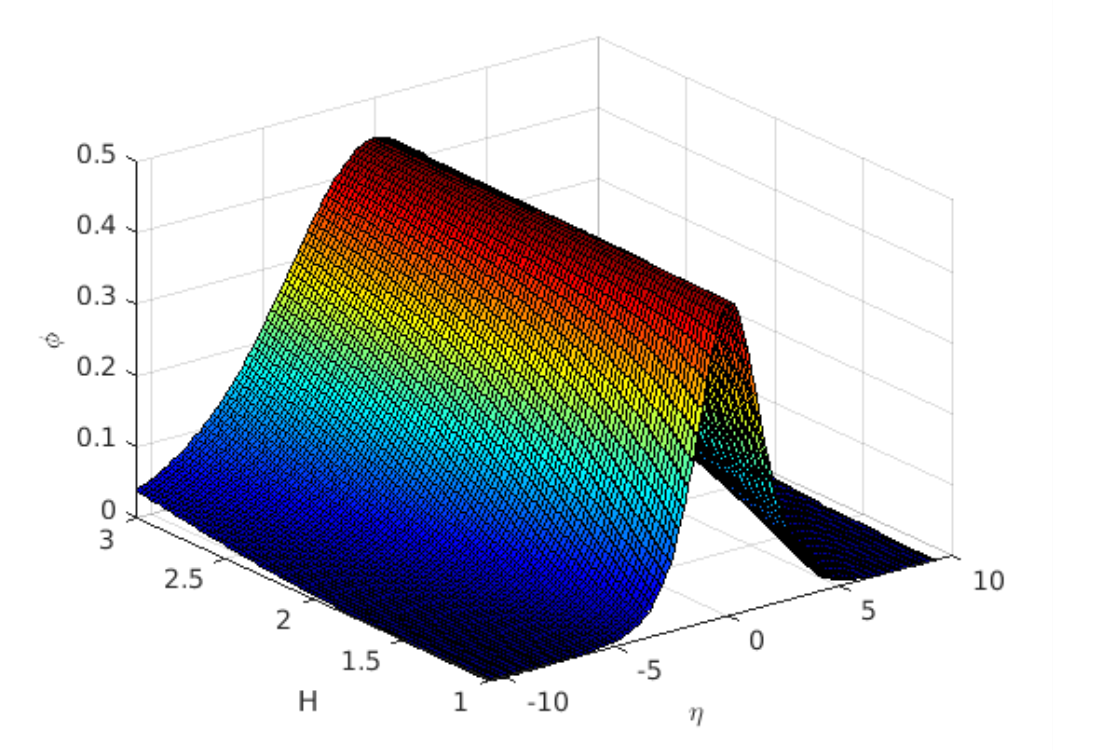}
	\caption{Solitary wave structure for different values of H}
	\label{fig4}
}
\end{figure}
\begin{figure}[!htb]
{	\centering
	\includegraphics[width=3in,angle=0]{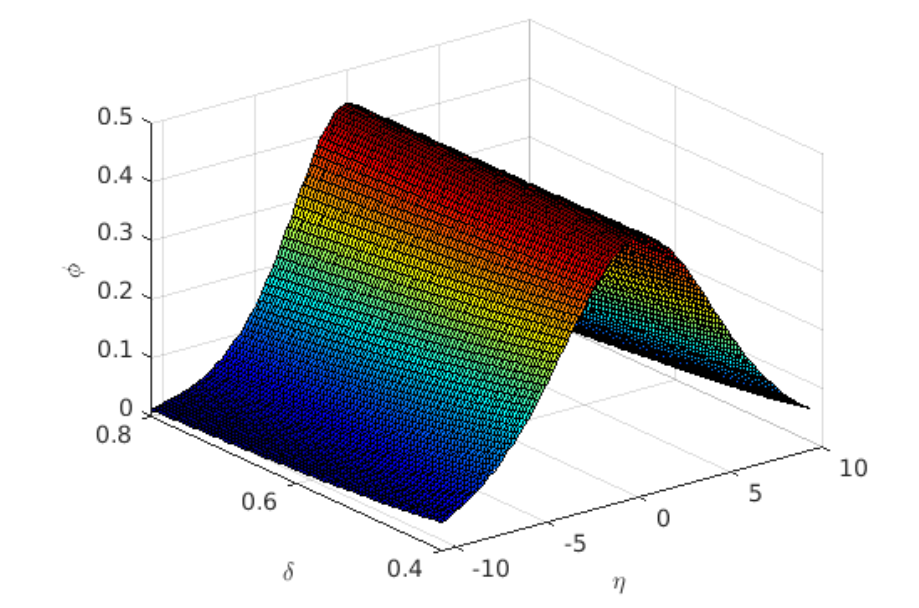}
	\caption{Solitary wave structure for different values of $\delta$}
	\label{fig5}
}
\end{figure}
\begin{figure}[!htb]
{	\centering
	\includegraphics[width=3in,angle=0]{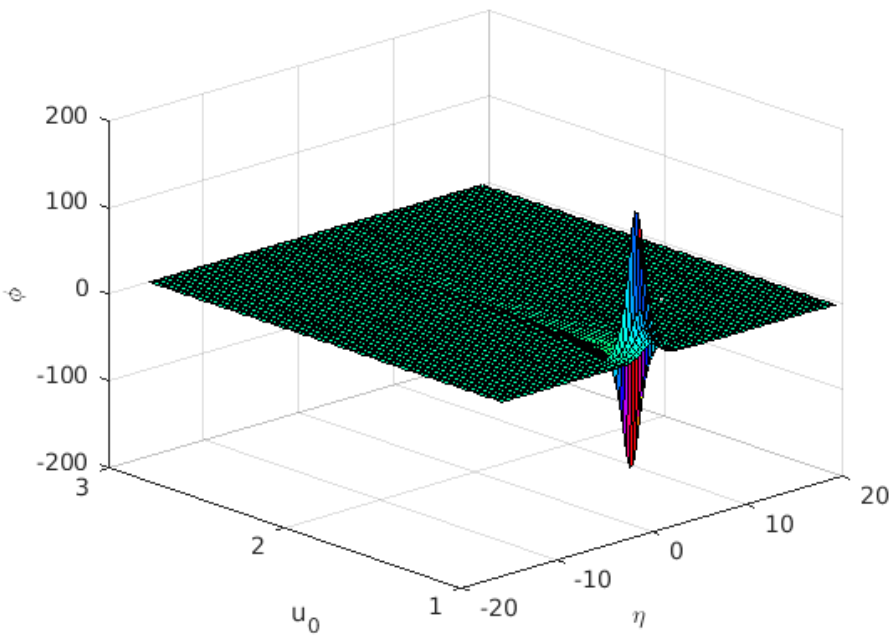}
	\caption{Solitary wave structure for different values of $u_0$}
	\label{fig6}
}
\end{figure}

What we see in Fig. (4), Fig. (5), and Fig. (6) are the three-dimensional representations of changing the structure of solitary wave structures. In Fig. (4) we see that with the increment of $H$ values the width of the structure increases. On the other hand in Fig. (5) the width of the structure decreases with increasing $\delta$ value. In Fig. (6), the picture is a little different. We see two peaks at two certain points i.e. for two particular values of streaming motion $u_0$ there is one maximum and one minimum in the potential for $u_0=1.09$ the potential shift is 161.5 in the positive side and for $u_0=1.12$ the potential shift is 133.6 in the negative side. To understand the changes in the amplitude we again plot these in two dimension planes.

\begin{figure}[!htb]
{	\centering
	\includegraphics[width=3in,angle=0]{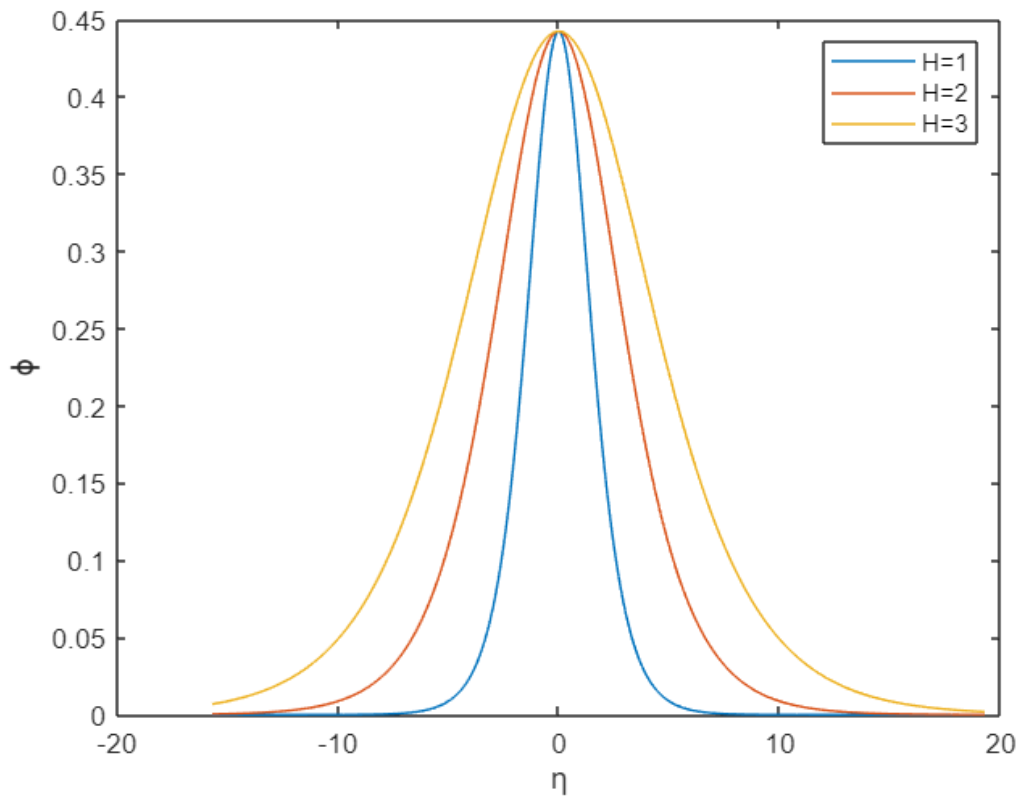}
	\caption{Solitary wave structure for different values of H}
	\label{fig7}
}
\end{figure}
\begin{figure}[!htb]
{	\centering
	\includegraphics[width=3in,angle=0]{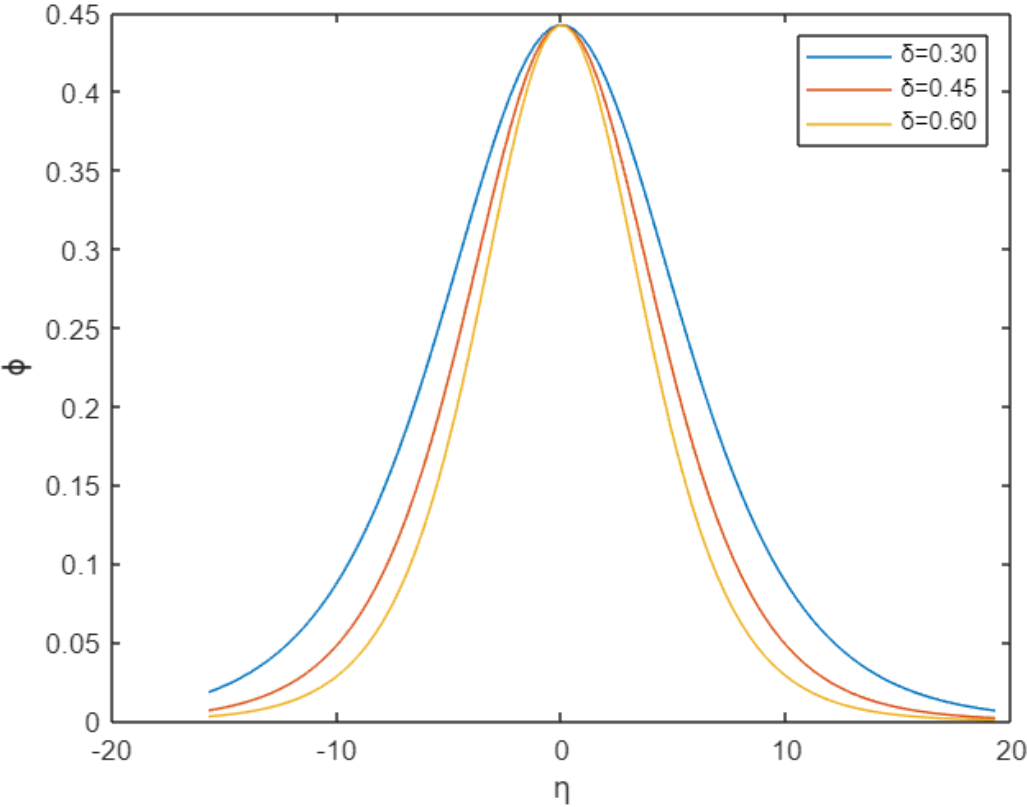}
	\caption{Solitary wave structure for different values of $\delta$}
	\label{fig8(a)}
}
\end{figure}
\begin{figure}[!htb]
{	\centering
	\includegraphics[width=3in,angle=0]{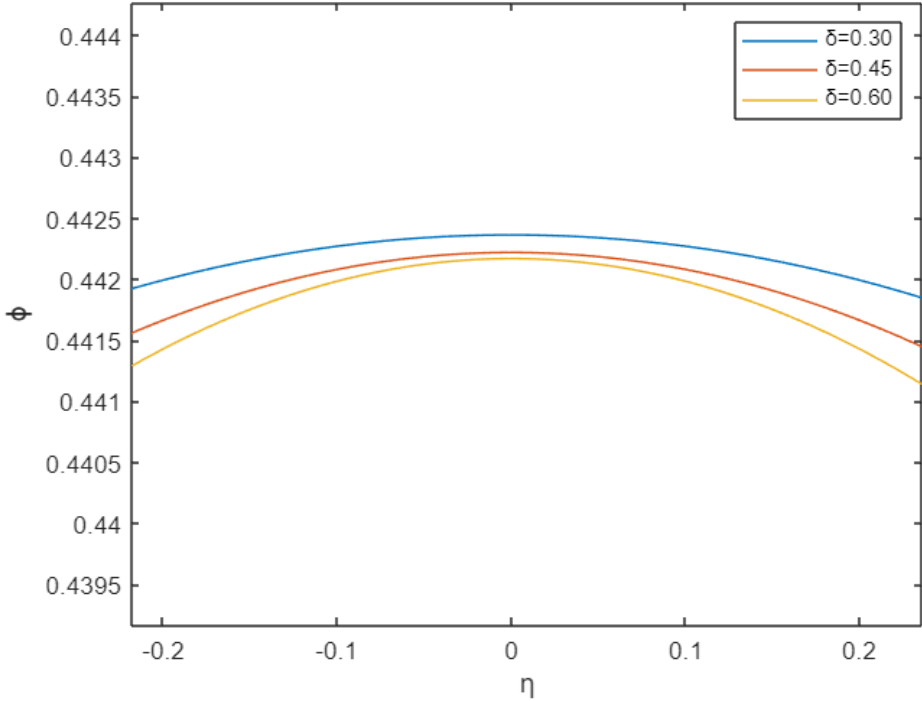}
	\caption{Zoomed solitary wave structure for different values of $\delta$}
	\label{fig8(b)}
}
\end{figure}
\begin{figure}[!htb]
{	\centering
	\includegraphics[width=3in,angle=0]{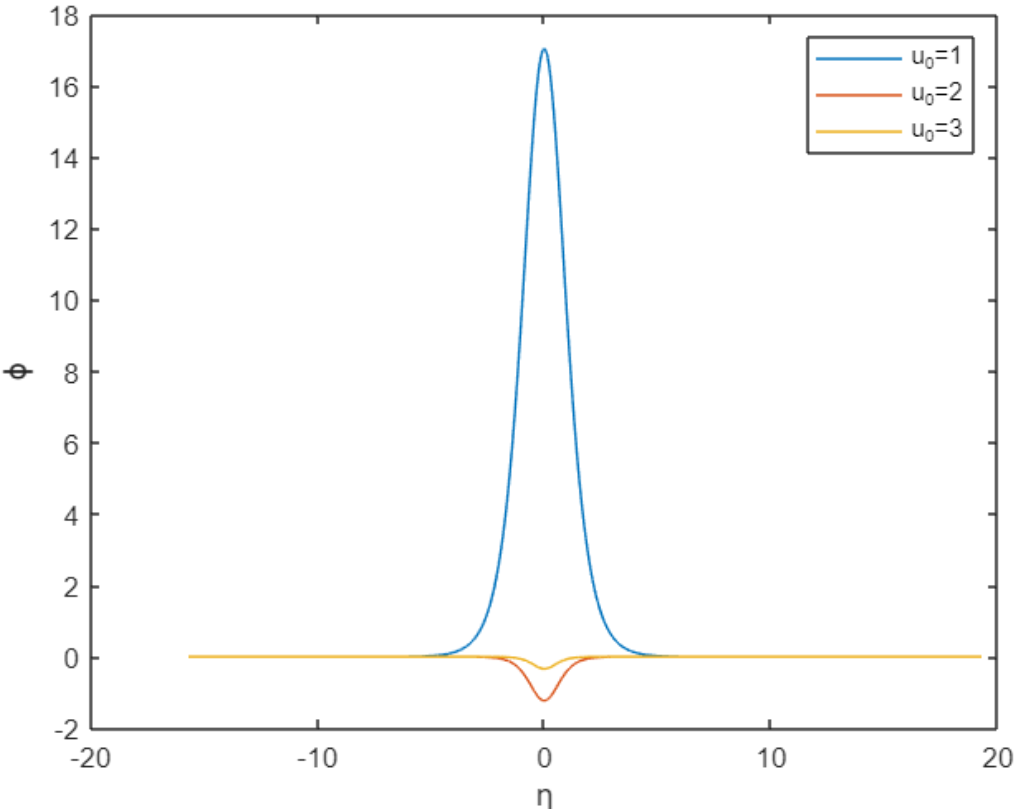}
	\caption{solitary wave structure for different values of $u_0$}
	\label{fig9}
}
\end{figure}

The amplitude of the structure is independent of $H$ values, whereas it depends very slightly with $\delta$ values. From Fig. (9) it can bee seen that the amplitude decreases slightly with increasing $\delta$ value. Fig. (10) represents the structures for different streaming motion. One can see that for some values of streaming motion the structure is giving rarefactive which peaks for $u_0=1.12$ and some values, it's giving compressive profile which peaks at $u_0=1.09$.
\section{Rogue wave}\label{RW}

Solitary waves are created due to the balance between the non-linear force and dispersive force in plasma, as the balance begins to break the KDV equation in Eq. (20) begins to modify to produce a sudden wave with a much higher amplitude than that of normal waves around at that time. These freak waves suddenly come to appear and disappear without any trace. We study the stability electron acoustic wave (EAW) and here we study that due to changes of some parameters modulation instability (MI) sets in and rogue waves come to the picture. The sudden hit of a rogue wave can create a huge disturbance. So, the study is important. It creates due to waves coming from different directions with different speeds interfering or it can be created due to the nonlinear effects of the plasma waves.
Any field variable can be expanded in terms of Fourier series as follows\cite{ref35}

\begin{equation}
    F=\epsilon^{2} F_{0}+\sum_{s=1}^{\infty} \epsilon_{s}\left(F_{s} e^{i s \psi}+F_{s}^{*} e^{-i s \psi}\right)
\end{equation}

Where,$F$ is the field variable, $\epsilon$ is the smallness parameter, and $\psi$ is the phase factor. Now, using this equation we can expand the field variable i.e. the potential in KDV equation

\begin{equation}
    \phi=\epsilon^{2} \phi_{0}+\epsilon \phi_{1} e^{i \psi}+\epsilon \phi_{1}^{*} e^{-i \psi}+\epsilon^{2} \phi_{2} e^{2 i \psi}+\epsilon^{2} \phi_{2}^{*} e^{-2 i \psi}+\cdots
\end{equation}

The first harmonics ($\phi_1$) and second harmonics ($\phi_2$) can be further expanded respectively

\begin{equation}
\phi_{1}=\phi_{1}^{(1)}+\epsilon \phi_{1}^{(2)}+\epsilon^{2} \phi_{1}^{(3)}+\ldots
\end{equation}
\begin{equation}
\phi_{2}=\phi_{2}^{(1)}+\epsilon \phi_{2}^{(2)}+\epsilon^{2} \phi_{2}^{(3)}+\ldots
\end{equation}

$F_0$ and $F_s$ are assumed to vary very slowly with space and time.
Now, we use the change in variables $\frac{\partial}{\partial \tau}=-i s \omega-\epsilon c \frac{\partial}{\partial \rho}+\epsilon^{2} \frac{\partial}{\partial \theta} \text { and } \frac{\partial}{\partial \xi}=i s k+\epsilon \frac{\partial}{\partial \rho}$ and put it in the Eq. (20) and equate the coefficients of the first order of $\epsilon$, we get

\begin{equation}
\omega=-B k^{3}
\end{equation}
\begin{equation}
\frac{d \omega}{d k}=-3 B k^{2}
\end{equation}
Equating the coefficients of $e^{2i\psi}$ with $\epsilon^2$
\begin{equation}
\varphi_{2}^{(1)}=\frac{A}{6 B k^{2}} \varphi_{1}^{(1)^{2}}
\end{equation}
Equating the terms independent of $\psi$ with $\epsilon^3$
\begin{equation}
    \varphi_{0}^{(1)}=\frac{A}{C} \varphi_{1}^{(1)} \varphi_{1}^{(1)^{*}}
\end{equation}
By equating the other higher terms we get
\begin{equation}
\frac{\partial \varphi_{1}^{(1)}}{\partial \tau}+3 i k B \frac{\partial^{2} \varphi_{1}^{(1)}}{\partial \rho^{2}}=-i A k\left(\varphi_{0}^{(1)} \varphi_{1}^{(1)}+\varphi_{2}^{(1)} \varphi_{1}^{(1)^{*}}\right)
\end{equation}
\begin{equation}
i \frac{\partial \varphi_{1}^{(1)}}{\partial \tau}-3 k B \frac{\partial^{2} \varphi_{1}^{(1)}}{\partial \rho^{2}}+\frac{A^{2}}{6 B k}\left(\varphi_{1}^{(1)^{2}} \varphi_{1}^{(1)^{*}}\right)=0
\end{equation}

Let’s assume $P= -3Bk$ and $Q= \frac{A^2}{6Bk}$ . The term {PQ} is very important in defining the existence of a rogue wave. In the region $PQ<0$ the rogue wave is stable and for $PQ>0$ the rogue wave doesn’t exist. Here, $P= -3Bk$ is always negative, hence $PQ= - A^2/2$ is always negative and we get rogue wave.
The solution of Eq. (41) we get as

\begin{equation}
    \phi(\rho, \theta)=\sqrt{\frac{2 P}{Q}}\left[\frac{4(1+4 i P Q)}{1+16 P^{2} Q^{2}+4 \rho^{2}}-1\right] \exp (2 i P \theta)
\end{equation}
Where, by changing the variables $\rho$ and $\theta$ we get the rogue wave structure for a very short range of those variables and the amplitude is much larger than that of normal waves.
\subsection{Result and discussions}
\begin{figure}[!htb]
{	\centering
	\includegraphics[width=3in,angle=0]{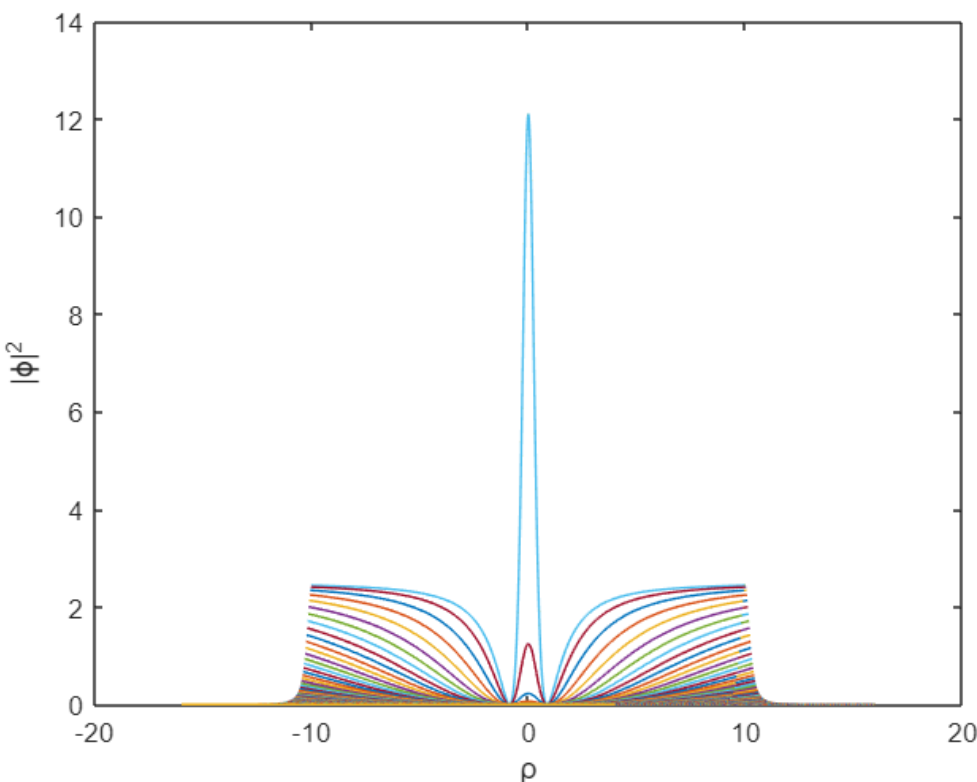}
	\caption{Rogue wave structure}
	\label{fig10}
}
\end{figure}
Here we plot the potential amplitude with the changed variable $\rho$ for different values of $\theta$. And we take almost 500 values of $\theta$ and as one can see that for only one value the potential suddenly gets abnormally high with respect to the other. All of the solutions we get from the same NLSE, where only one wave behaves dangerously. It perfectly depicts the feature of a Rogue Wave, the amplitude is way higher than that of solitary profiles in Fig. (7), Fig. (8), and Fig. (9). 
\section{Simulation}
Here, in this section, we show the static plots of two simulations of the solitary profiles and rogue wave profile of the relativistic plasma we are working on. These two simulations were run in MATLAB.

\begin{figure}[!htb]
{	\centering
	\includegraphics[width=3in,angle=0]{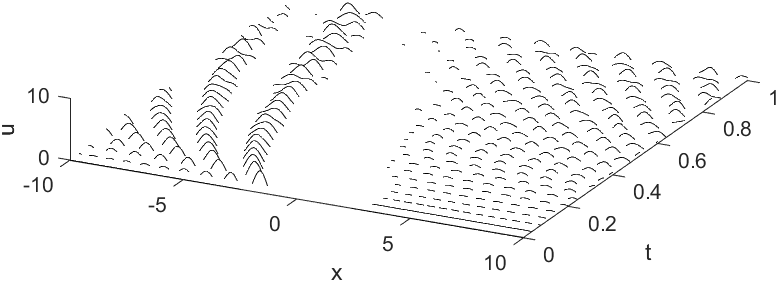}
	\caption{Simulation for KDV solitary profile}
	\label{fig10}
}
\end{figure}
\begin{figure}[!htb]
{	\centering
	\includegraphics[width=3in,angle=0]{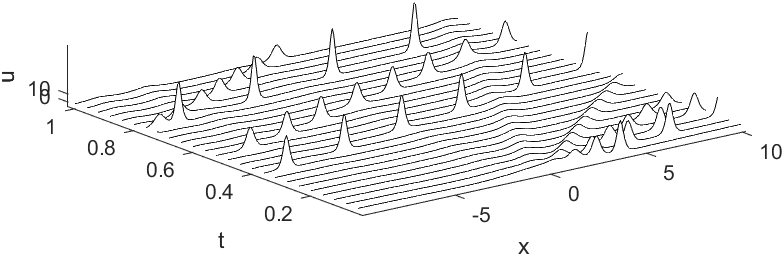}
	\caption{Simulation for rogue wave}
	\label{fig10}
}
\end{figure}
In both figures, using the solutions of Eq. (20) and Eq. (43) we simulate the KDV solitary wave and rogue wave respectively. Here, ‘x’ represents the space coordinate and ‘t’ represents the time coordinate, and ‘u’ is the amplitude. In both cases, we see the changes in the waves in the interval between t = 0s and t = 1s. As you can see from the simulations that rogue waves have much higher amplitudes than those of normal solitary waves and also less frequent than normal waves.
\section{Conclusions}\label{Conc}
Our whole work surrounds the nonlinear effects of one-dimensional electron acoustic waves (EAW) in plasma. In various situations regarding the earth’s atmosphere, laboratory plasma, or space, EAW is of great interest to the scientists. People have worked with ion-acoustic waves under Fermi pressure and also electron acoustic waves under relativistic degeneracy pressure. Here, we work with the electron-acoustic wave of quantum plasma under Fermi pressure. Though the motion of cold electrons is much lower than that of hot electrons here we consider the relativistic motion for both species of electrons and consider ions to be static due to their heavy mass with respected to the electrons. We study the dispersion relations of the wave for various values of streaming motion, diffraction coefficient, and the ratio of hot to cold electrons. It can be seen that the higher the value of quantum diffraction term and streaming velocity the higher is the nonlinear effect. On the other hand, nonlinear effects increase when one of the hot and cold electron species gets much higher in numbers than others. We study the solitary wave structure using the KDV equation and the plots show the changes in the wave amplitude and width with the changing value of diffraction term and ratio term. Then we show from the Non-linear Schrödinger equation stable rogue wave is forming. The wave’s amplitude is way higher than that of normal waves. At last through fine simulations, the evolution of solitary structure and rogue waves in time can be seen. Hence, the detailed study of EAW in plasma under Fermi pressure with the relativistic motion of the electrons with the plots and simulations may help to understand the situations in the super dense cores of white dwarfs, compact stars with considering weakly relativistic regime.

\bibliographystyle{IEEEtran}
\bibliography{references.bib}
\cleardoublepage

\vfill
\vfill
\vfill
\vfill
\vfill
\vfill
\vfill
\vfill
\vfill
\vfill
\vfill
\vfill
\vfill
\vfill
\vfill
\vfill
\vfill
\vfill
\vfill
\vfill
\vfill
\vfill
\vfill
\vfill
\vfill
\vfill
\vfill
\vfill
\vfill
\vfill
\vfill
\vfill
\vfill
\vfill
\vfill
\vfill
\vfill
\vfill
\vfill
\vfill
\vfill
\vfill
\vfill
\vfill
\end{document}